\documentclass{article}
\usepackage[preprint]{spconf}
\usepackage{amsmath,graphicx,amssymb,pifont}
\usepackage{booktabs,multirow,multicol,fancyhdr}
\usepackage[colorlinks=True]{hyperref}
\usepackage[numbers,sort&compress]{natbib}
\usepackage{soul}

\copyrightnotice{\copyright\ IEEE 2022}
\toappear{To appear in {\it Proc.\ ICASSP2022, May 22-27, 2022, Singapore}}

\usepackage{xcolor}
\usepackage{tipa}

\title{Improving the fusion of acoustic and text representations in RNN-T}

\name{Chao Zhang, Bo Li, Zhiyun Lu, Tara N. Sainath and Shuo-yiin Chang\thanks{Thanks Dr. Zhehuai Chen and Dr. Matt Shannon for useful suggestions.}}
\address{Google LLC, USA \\
\fontsize{9}{9}\selectfont\ttfamily\upshape
\{chaoz, boboli, zhiyunlu, tsainath, shuoyiin\}@google.com}
\begin{document}
\ninept
\maketitle
\begin{abstract}
The recurrent neural network transducer (RNN-T) has recently become the mainstream end-to-end approach for streaming automatic speech recognition (ASR). To estimate the output distributions over subword units, RNN-T uses a fully connected layer as the joint network to fuse the acoustic representations extracted using the acoustic encoder with the text representations obtained using the prediction network based on the previous subword units. In this paper, we propose to use gating, bilinear pooling, and a combination of them in the joint network to produce more expressive representations to feed into the output layer. A regularisation method is also proposed to enable better acoustic encoder training by reducing the gradients back-propagated into the prediction network at the beginning of RNN-T training. Experimental results on a multilingual ASR setting for voice search over nine languages show that the joint use of the proposed methods can result in 4\%--5\% relative word error rate reductions with only a few million extra parameters. 
\end{abstract}
\section{Introduction}
\label{sec:intro}
In contrast to the traditional modular-based ASR system that consists of an acoustic model, a language model (LM) and a rule-based decoder,  
the recent end-to-end (E2E) approach aims at implementing the ASR process using a single neural network model. In particular, both attention-based encoder-decoder \cite{Chorowski2015AttentionBasedMF,Lu2015ASO,Chan2016ListenAA,Kim2017} and recurrent neural network transducer (RNN-T) \cite{Graves2012,Graves2013,Prabhavalkar2017,Battenberg2017,JinyuLi2019,Zeyer2020} methods achieve a 
coherently integrate acoustic and text information
using a recurrent structure between the previous and current subword units in the output text sequence  
in contrast to
modular-based ASR systems.  Recently, RNN-T has become more prevalent due to its streaming benefits \cite{He2019,Chang2019,BoLi2020,QianZhang2020,Le2021,Shi2021,XieChen2021}. 

RNN-T was first proposed to extend a connectionist temporal classification (CTC) \textit{acoustic encoder} \cite{Graves2006} with a \textit{prediction network} serving as an LM. Acoustic and text representations over phonemes are derived separately from the acoustic encoder and prediction network, and fused using an addition followed by a softmax function to produce the final output distributions \cite{Graves2012}. Shortly afterward, an improvement was introduced to fuse more compressed hidden representations using concatenation and a fully connected (FC) layer with a hyperbolic tangent (tanh) function \cite{Graves2013}, which is termed as the \textit{joint network}. The fused representations are further transformed by the output layer, which is another FC layer with the softmax function, into the output distributions.
Since then, there has been a research focus to improve the RNN-T encoder structure, from using the long short-term memory (LSTM) model \cite{Hochreiter1997,Rao2017} to Transformer and Conformer \cite{QianZhang2020,Gulati2020}, and to improve their streaming performance and on-device efficiency \cite{He2019,Narayanan2021,YuekaiZhang2021}.
More recently, many studies focus on improving the recognition accuracy on long tail words/phrases and long-form utterances, which results in the use of extra model components \cite{Le2021,BrianSun2021tcp}, novel prediction network structures and decoding algorithms \cite{Prabhavalkar2017,Ghodsi2020,Saon2020}, 
alternative subword units to output \cite{Rao2017,He2019},
test-time external language model integration \cite{McDermott2019,Variani2020,Meng2021}, and
synthetic data augmentation and knowledge distillation methods \cite{He2019,JinyuLi2020,TonyZheng2021,Doutre2021}.

Though joint network is the component closest to the output layer and is important to RNN-T performance, there are only a few studies related to it \cite{Graves2013,Variani2020,Saon2021}. 
In this paper, by viewing the function of the joint network as to fuse the representations of the acoustic and text modalities, we propose to improve the joint network implementation using different structures for information fusion, including gating and a low-rank approximation of bilinear pooling with shortcut connections and a tanh transform. A better-performing structure is further proposed by stacking bilinear pooling on top of gating.  
Furthermore, since text priors are often easier to learn than complex acoustic patterns, the prediction network often converges much faster than the acoustic encoder that causes the joint network overly biased towards the prediction network. To alleviate this issue, a novel regularisation method is proposed to penalise the gradients back-propagated into the prediction network in early training stages, which can improve RNN-T performance without bringing any cost to both training and test. Experiments were conducted on a large-scale multilingual ASR setup for voice search, similar to the one used in \cite{BoLi2021}. As a result, by jointly using these proposed methods, more than 4\% relative word error rate (WER) reductions were achieved by only increasing a few million extra parameters.

The rest of the paper is organised as follows: Sec.~\ref{sec:background} reviews RNN-T and the work related to joint network. Sec.~\ref{sec:approach} gives details of the proposed joint network structures and the regularisation method. Sec.~\ref{sec:setup} describes our experimental setup, followed by the discussions on the results in Sec.~\ref{sec:results}. We conclude in Sec.~\ref{sec:conclusions}.

\section{Background}
\label{sec:background}

\subsection{RNN Transducer}
\label{ssec:rnnt}
In the traditional statistical ASR framework, 
\cite{jelinek1997statistical},
speech is produced and encoded via a noisy channel and the ASR system is to find the most probable source text sequence $\mathbf{y}^{*}$ given the acoustic feature sequence $\mathbf{x}_{1:T}$ of length $T$ observed as the output of the channel. Based on Bayes' rule, decoding follows the \textit{maximum a posteriori} rule to search over each possible hypothesized text sequence $\mathbf{y}$ by
\begin{equation}
    P(\mathbf{y}|\mathbf{x}_{1:T})\propto p(\mathbf{x}_{1:T}|\mathbf{y})P(\mathbf{y}),
\end{equation}
where $p(\mathbf{x}_{1:T}|\mathbf{y})$ is estimated by the acoustic model and is the likelihood of generating $\mathbf{x}_{1:T}$ through the channel; $P(\mathbf{y})$ is estimated by an LM, 
describing the underlying probabilistic distribution of the source text.

Instead of modelling $p(\mathbf{x}_{1:T}|\mathbf{y})$ and $P(\mathbf{y})$ by independent models in a modularised system, E2E methods, such as RNN-T, directly models $P(\mathbf{y}|\mathbf{x}_{1:T})$ by a single model.
Let $\mathbf{y}={y}_{1:U}$ where $U$ is the number of subword units in $\mathbf{y}$. For a streaming setting without any look ahead frame and time reduction, $\mathbf{h}^{\text{enc}}_{t}$, the $D^{\text{enc}}$-dimensional (-dim) acoustic representation extracted by the acoustic encoder at time $t$, $\mathbf{h}^{\text{pred}}_{u}$, the $D^{\text{pred}}$-dim text representation of the $u$-th subword unit by the prediction network, and $\mathbf{h}^{\text{joint}}_{t,u}$, the $D^{\text{joint}}$-dim fused representation generated by the joint network, are calculated as follows:
\begin{align}
    &\label{eq:encoder}\mathbf{h}^{\text{enc}}_{t}=\text{AcousticEncoder}(\mathbf{x}_{1:t}),\\
    &\label{eq:predictor}\mathbf{h}^{\text{pred}}_{u}=\text{PredictionNetwork}(y_{1:u-1}),\\
    &\label{eq:fusion}\mathbf{h}^{\text{joint}}_{t,u}=\text{JointNetwork}(\mathbf{h}^{\text{enc}}_{t},\mathbf{h}^{\text{pred}}_{u}),\\
    &\label{eq:output}P(\hat{y}_i=k|y_{0:u-1},\mathbf{x}_{1:t})=\text{Softmax}(\text{W}^{\text{out}}\mathbf{h}^{\text{joint}}_{t,u})|_{k},
\end{align}
where $y_0$ refers to the special start of sentence symbol; $k$ and $\mathbf{W}^{\text{out}}$ are the $k$-th node and weights of the output layer. Regarding a set of subword units $\mathcal{V}$, the symbol that $k$ represents belongs to $\mathcal{V}\cup\{\varnothing\}$, where $\varnothing$ is the blank symbol indicating no subword is emitted. 
During training, let $\hat{\mathbf{y}}=\{\hat{y}_1,\hat{y}_2,\ldots,\hat{y}_{T+U}\}$ be an alignment sequence of $\mathbf{y}$ that can be converted into $\mathbf{y}$ by removing all occurrences of $\varnothing$, $\mathcal{A}(\mathbf{x}_{1:T},\mathbf{y})$ be the reference lattice including all possible alignment sequences between $\mathbf{y}$ and $\mathbf{x}_{1:T}$, $P(\mathbf{y}|\mathbf{x}_{1:T})$ can be computed efficiently with the \textit{forward-backward procedure}. There is
\begin{align}
P(\mathbf{y}|\mathbf{x}_{1:T})=\sum\nolimits_{\hat{\mathbf{y}}\in\mathcal{A}(\mathbf{x}_{1:T},\mathbf{y})}\prod\nolimits^{T+U}_{i=1}P(\hat{y}_i|\mathbf{x}_{1:t_i},y_{1:u_i}),
\end{align}
where $t_i$ and $u_i$ are the values of $t$ and $u$ corresponding to $\hat{y}_i$ in $\hat{\mathbf{y}}$. 

In practice, $\text{AcousticEncoder}(\cdot)$ can be Conformer with a fixed number of look ahead frames and a fixed time reduction rate. $\text{PredictionNetwork}(\cdot)$ is often a multi-layer LSTM. The joint network is often defined as 
an FC layer \cite{Graves2013} that
\begin{align}
\label{eq:fusion_fc}
\mathbf{h}^{\text{joint}}_{t,u}=\tanh(\mathbf{W}^{\text{joint}}_{1}\mathbf{h}^{\text{enc}}_{t}+\mathbf{W}^{\text{joint}}_{2}\mathbf{h}^{\text{pred}}_{u}),
\end{align}
where $\mathbf{W}^{\text{joint}}_1$ and $\mathbf{W}^{\text{joint}}_2$ are weight matrices.
For simplicity, bias is ignored in Eqn.~\eqref{eq:fusion_fc} and the rest of the paper.

If $\mathbf{W}^{\text{joint}}_2$ is $\mathbf{0}_{D^{\text{joint}}\times D^{\text{pred}}}$ and the joint network transforms only the acoustic representation, apart from their difference in $\mathcal{A}(\mathbf{x}_{1:T},\mathbf{y})$ \cite{He2019,Variani2020}, RNN-T becomes CTC that calculates $P(\hat{y}_i|\mathbf{x}_{1:t_i})$ by making an independence assumption between any subword units in $\mathbf{y}$.
This reveals the importance of the joint network.


\subsection{Related work}
\label{ssec:relatedwork}
In Eqn.~\eqref{eq:fusion_fc}, by enforcing $\mathbf{h}^{\text{enc}}_{t}=\mathbf{0}$, the prediction network, joint network, and output layer jointly form an LSTM LM that is often referred to as the \textit{internal LM} \cite{Variani2020}. 
Studies showed that more WER reductions can be found from  shallow fusion with external LMs by discounting the internal LM scores at test-time \cite{McDermott2019,Variani2020,Meng2021}. 
More recently, stateless RNN-T has been proposed to truncate the LM history embedded in the prediction network to $n$-gram \cite{Ghodsi2020}.  

The fusion of representations associated with different modalities plays a key role in multimodal intelligence \cite{ChaoZhang2020}: attention, gating, and bilinear pooling are the most commonly used structures for the purpose. In RNN-T, the standard joint network implemented based on Eqn.~\eqref{eq:fusion_fc} can be viewed as to fuse the acoustic representation $\mathbf{h}^{\text{enc}}_{t}$ and text representation $\mathbf{h}^{\text{pred}}_{u}$ using an FC layer. Recently, an alternative joint network structure is proposed to model the multiplicative interactions between the two representations \cite{Saon2021}:
\begin{align}
\label{eq:fusion_mul}
\mathbf{h}^{\text{joint}}_{t,u}=\tanh(\mathbf{W}^{\text{joint}}_{1}\mathbf{h}^{\text{enc}}_{t}\odot\mathbf{W}^{\text{joint}}_{2}\mathbf{h}^{\text{pred}}_{u}),
\end{align}
where $\odot$ is the Hadamard product. 

\section{Fusing acoustic and text representations}
\label{sec:approach}
Fusing acoustic and text representations is arguably a difficult task, and the standard joint network simply uses one FC layer. Sec.~\ref{ssec:gating} and \ref{ssec:bilinear} propose more complex joint network structures, and Sec.~\ref{ssec:internallm} proposes to improve the balance between the two modalities.

\subsection{Gating mechanism}
\label{ssec:gating}
Gating has been widely used in recurrent and shortcut structures \cite{Hochreiter1997,YuZhang2016}, 
whose most famous application is LSTM. It allows each element in each representation vector to be scaled with a different dynamic weight, before being integrated via vector addition. Specifically,
\begin{align}
\label{eq:fusion_gate}
\mathbf{g}_{t,u}&=\sigma(\mathbf{W}^{\text{gate}}_{1}\mathbf{h}^{\text{enc}}_{t}+\mathbf{W}^{\text{gate}}_{2}\mathbf{h}^{\text{pred}}_{u}),\\
\nonumber\mathbf{h}^{\text{joint}}_{t,u}&=\mathbf{g}_{t,u}\odot\tanh(\mathbf{W}^{\text{joint}}_{1}\mathbf{h}^{\text{enc}}_{t})+(1-\mathbf{g}_{t,u})\odot\tanh(\mathbf{W}^{\text{joint}}_{2}\mathbf{h}^{\text{text}}_{u}),
\end{align}
where $\mathbf{g}_{t,u}$ 
is the gating vector, $\sigma(\cdot)$ is sigmoid function, and $\mathbf{W}^{\text{gate}}_{1}$ and $\mathbf{W}^{\text{gate}}_{2}$ are 
weight matrices of the gating layer. Notably a different gating vector can be computed to replace $1-\mathbf{g}_{t,u}$. However, we observed worse WERs when using two separate gating vectors.

\subsection{Bilinear pooling}
\label{ssec:bilinear}
Compared to gating, bilinear pooling is a more powerful and expensive method for fusing multimodal representations \cite{Tenenbaum2000}, which combines two vectors using the bilinear form (with bias ignored)
\begin{align}
\label{eq:fusion_bl1}
{h}^{\text{joint}}_{t,u,d}=(\mathbf{h}^{\text{enc}}_{t})^{\text{T}}\mathbf{W}^{\text{bi}}_{d}\,\mathbf{h}^{\text{pred}}_{u},
\end{align}
where 
$\mathbf{W}^{\text{bi}}_{d}$ is a $D^{\text{enc}}\times D^{\text{pred}}$-dim matrix, and ${h}^{\text{joint}}_{t,u,d}$ is the $d$-th element of $\mathbf{h}^{\text{joint}}_{t,u}$.
Considering all elements in $\mathbf{h}^{\text{joint}}_{t,u}$,
 $[\mathbf{W}^{\text{bi}}_{1},\ldots,\mathbf{W}^{\text{bi}}_{D^{\text{joint}}}]$
is a $D^{\text{enc}}\times D^{\text{pred}}\times D^{\text{joint}}$-dim weight tensor. Fusing $\mathbf{h}^{\text{enc}}_{t}$ and $\mathbf{h}^{\text{pred}}_{u}$ into $\mathbf{h}^{\text{joint}}_{t,u}$ is therefore equivalent to perform
\begin{align*}
\mathbf{h}^{\text{joint}}_{t,u}=[\text{Vector}(\mathbf{W}^{\text{bi}}_{1}),\ldots,\text{Vector}(\mathbf{W}^{\text{bi}}_{D^{\text{joint}}})]^{\text{T}}\text{Vector}(\mathbf{h}^{\text{enc}}_{t}\otimes\mathbf{h}^{\text{pred}}_{u}),
\end{align*}
where $\text{Vector}(\cdot)$ and $\otimes$ are the vectorisation and outer product operations. Compared to gating, bilinear pooling first computes the outer product of the two vectors to capture the multiplicative interactions between all possible element pairs in a more expressive  $D^{\text{enc}}\times D^{\text{pred}}$-dim space, and then projects it into a $D^{\text{joint}}$-dim vector space. 

To avoid the issues when estimating a high dimensional weight tensor,  
a low-rank approximation $\mathbf{W}^{\text{bi}}_{d}\approx\mathbf{W}^{\text{low}}_{1,d}(\mathbf{W}^{\text{low}}_{2,d})^{\text T}$ is suggested \cite{Pirsiavash2009}, where $\mathbf{W}^{\text{low}}_{1,d}$ and $\mathbf{W}^{\text{low}}_{2,d}$ are $D^{\text{enc}}\times D^{\text{rank}}$-dim and $D^{\text{pred}}\times D^{\text{rank}}$-dim matrices, and $D^{\text{rank}}$ is the rank of $\mathbf{W}^{\text{bi}}_{d}$. Therefore Eqn.~\eqref{eq:fusion_bl1} can be rewritten as ${h}^{\text{join}}_{t,u,d}\approx(\mathbf{h}^{\text{enc}}_{t})^{\text{T}}\mathbf{W}^{\text{low}}_{1,d}(\mathbf{W}^{\text{low}}_{2,d})^{\text T}\mathbf{h}^{\text{pred}}_{u}=\mathbf{1}^{\text T}((\mathbf{W}^{\text{low}}_{1,d})^{\text T}\mathbf{h}^{\text{enc}}_{t}\odot(\mathbf{W}^{\text{low}}_{2,d})^{\text T}\mathbf{h}^{\text{pred}}_{u})$. It was proposed to tie all $\mathbf{W}^{\text{low}}_{1,d}$ as $\mathbf{W}^{\text{low}}_{1}$ and all $\mathbf{W}^{\text{low}}_{2,d}$ as $\mathbf{W}^{\text{low}}_{2}$, and to use a projection matrix $\mathbf{W}^{\text{proj}}$ to distinguish the elements in $\mathbf{h}^{\text{joint}}_{t,u}$ \cite{JHKim2017}. 
When a tanh function is used to transform the vectors before the Hadamard product, there is 
\begin{align}
\label{eq:fusion_bl3}
    \hat{\mathbf{h}}^{\text{joint}}_{t,u}=\mathbf{W}^{\text{proj}}(\tanh((\mathbf{W}^{\text{low}}_{1})^{\text T}\mathbf{h}^{\text{enc}}_{t})\odot\tanh((\mathbf{W}^{\text{low}}_{2})^{\text T}\mathbf{h}^{\text{pred}}_{u})).
\end{align}
We found using shortcut connections \cite{BrianSun2021speaker} and a final tanh transform are important for bilinear pooling for RNN-T, and thus propose
\begin{align}
\label{eq:fusion_bl4}
{\mathbf{h}}^{\text{joint}}_{t,u}=\tanh(\hat{\mathbf{h}}^{\text{joint}}_{t,u}+\mathbf{W}^{\text{joint}}_1\mathbf{h}^{\text{enc}}_{t}+\mathbf{W}^{\text{joint}}_2\mathbf{h}^{\text{pred}}_{u}),
\end{align}
where $\mathbf{W}^{\text{joint}}_1\mathbf{h}^{\text{enc}}_{t}$ and $\mathbf{W}^{\text{joint}}_2\mathbf{h}^{\text{pred}}_{u}$ are the shortcut connections here.

At last, we propose a stack structure to combine gating and bilinear pooling to leverage their complementarity. It is implemented by replacing Eqn.~\eqref{eq:fusion_bl3} with 
\begin{align}
\label{eq:fusion_bl5}
    \hat{\mathbf{h}}^{\text{joint}}_{t,u}=\mathbf{W}^{\text{proj}}(\tanh((\mathbf{W}^{\text{low}}_{1})^{\text T}\mathbf{h}^{\text{enc}}_{t})\odot\tanh((\mathbf{W}^{\text{low}}_{2})^{\text T}\mathbf{h}^{\text{gate}}_{t,u})), 
\end{align}
where $\mathbf{h}^{\text{gate}}_{t,u}$ refers to the joint representation computed by Eqn.~\eqref{eq:fusion_gate}.

\subsection{Prediction network regularisation}
\label{ssec:internallm}
It has been observed that strong and prevalent text priors (\textit{e.g.} ``bananas are yellow'') often caused image-text multimodal systems to overfit to the statistical biases and tendencies, and largely circumvents the need to understand visual scenes \cite{Ramakrishnan2018}. 
Similary in RNN-T, since text priors are often easier to learn than the acoustic patterns, it is possible that the faster converging speed of the prediction network (than the acoustic encoder)
makes $\mathbf{h}^{\text{pred}}_{u}$ 
overly weighted in $\mathbf{h}^{\text{joint}}_{u,t}$. In that situation, the acoustic encoder is less well trained to handle the audio samples with high internal LM scores. 

In order to resolve this issue, we propose a prediction network regularisation method applied to the beginning of RNN-T training. It is implemented by recomputing the text representation as
\begin{align}
\label{eq:regu1}
    \mathbf{h}^{\text{pred}}_{u}=\alpha_m\,\mathbf{h}^{\text{pred}}_{u}-\text{sg}((\alpha_m-1)\,\mathbf{h}^{\text{pred}}_{u}),
\end{align}
where $m$ is the index of current training step, 
$\alpha_m$ is a scaling factor, 
$\text{sg}(\cdot)$ is the stop gradient function whose input tensor will have zero gradients. When  $0\leqslant\alpha_m\leqslant1$, the value of $\mathbf{h}^{\text{pred}}_{u}$ will not be changed but its corresponding gradients that back-propagate into the prediction network will be reduced by a factor of $\alpha_m$. This slows down the convergence of the prediction network and makes the joint network fuse $\mathbf{h}^{\text{enc}}_{t}$ and $\mathbf{h}^{\text{pred}}_{u}$ in a more balanced way. In this paper, a piece-wise linear scheduler is used to control the value of $\alpha_m$:
\begin{align}
\label{eq:regu2}
    \alpha_m=\left\{
    \begin{array}{lll} 0 & \text{if } m<m_1 \\
    1 & \text{else if } m\geqslant m_2\\
    (m-m_1) / (m_2-m_1) & \text{otherwise}   \\
    \end{array}\right.,
\end{align}
where $m_1$ and $m_2$ are two pre-defined hyper-parameters. 
Notably, this method is different from initialising RNN-T with a pre-trained CTC model even when $\alpha_m=0$, since the prediction network serves as a random but fix-valued projection, through which RNN-T is still able to obtain $y_{u-1}$. This links to the stateless RNN-T \cite{Ghodsi2020}.    

\section{Experimental Setup}
\label{sec:setup}
\subsection{Datasets}
\label{ssec:setup1}
Experiments were conducted on a dataset with 9 language locales: US English (en-US), UK English (en-GB), French (fr-FR), Italian (it-IT), Germany (de-DE), US Spanish (es-US), ES Spanish (es-ES), Taiwan Chinese (zh-TW) and Japanese (ja-JP). All data are anonymised and hand-transcribed. There are totally 214.2M utterances which correspond to 142.3K hours of speech data collected from Google’s Voice Search traffic. en-US and en-GB take about 25\% and 5\% of the training data, while each of the rest 7 languages takes about 10\% of the data. The SpecAugment method is used to improve ASR robustness against noisy conditions \cite{Park2020SpecAug}. The training data is mixed with all languages without using any language id information. The test sets are kept distinct for each language with each of them containing  3.3K$\sim$15.4K utterances. 
The testing utterances are also sampled from Google’s Voice Search traffic with a maximum duration constraint of 5.5 second long for each utterance. The test sets have no overlapping from the training set for evaluation purpose. 

\subsection{Model setup}
\label{ssec:setup2}
The 80-dim log-mel filter bank features are used, which are computed using a 32ms frame length and a 10ms shift. Acoustic features from 3 contiguous frames are stacked and
subsampled to form a 240-dim input representation with 30ms frame rate, which are then transformed using a linear projection to 512-dim and added with positional embeddings. Twelve Conformer \cite{Gulati2020} encoder blocks with 8-head self-attention and a convolution kernel size of 15 are used to further transform 
the stacked features. 
A concatenation operation is performed after the 3rd block to achieve a time reduction rate of 2, and the resulted 1024-dim vectors are transformed by the 4th Conformer block and then projected back to 512-dim using another linear transform. Afterwards comes with another 8 Conformer blocks followed by a final linear normalisation layer. These layers combined together form the RNN-T acoustic encoder. The prediction network consists of two layers of 2,048-dim LSTM with a 640-dim linear projection to make $D^{\text{pred}}=$640. The dimension of the fused representation $D^{\text{joint}}$ is also set to 640. 
All models are trained to predict 16,384 word-piece units \cite{Schuster2012}. 
As a result, the final RNN-T baseline has 110M parameters in the acoustic encoder and 33M parameters in the rest of the model.
All models are trained in Tensorflow using the Lingvo toolkit \cite{Shen2019Lingvo} on Google’s Tensor Processing Units V3 with a global batch size of 4,096 utterances. Models are optimized using synchronized stochastic gradient descent based on the Adam optimiser with $\beta_1=0.9$ and $\beta_2=0.999$. 
During test, the models are tested in a fully E2E fashion without any external LMs.

\section{Experimental Results}
\label{sec:results}
In this section, RNN-T systems with different joint network structures are first compared at the 200K-th training step, whose details are listed in Table~\ref{tab:models}. Next, prediction network regularisation with different hyper-parameter values are compared using S4 at the 500K-th step. Final results are presented with 800K training steps. 

\begin{table}[!htbp]
    \caption{Details of systems with different joint network structures. ``\#Params'' refers to the number of joint network parameters.}
    \centering
    \begin{tabular}{llcccc}
        \toprule
        ID & Structure & Equations & $D^{\text{joint}}$ & $D^{\text{rank}}$ & \#Params\\
        \midrule
        S0 & FC with add. & \eqref{eq:fusion_fc} & 640 & -- & 0.73M \\
        S1 & FC with add. & \eqref{eq:fusion_fc} & 790 & -- & 3.36M \\
        S2 & FC with mul. & \eqref{eq:fusion_mul} & 640 & -- & 0.73M \\
        \midrule
        S3 & Gating & \eqref{eq:fusion_gate} & 640 & -- & 1.47M \\
        S4 & Bilinear& \eqref{eq:fusion_bl3} \& \eqref{eq:fusion_bl4} & 640 & 640 & 1.88M \\
        S5 & Bilinear & \eqref{eq:fusion_bl3} \& \eqref{eq:fusion_bl4} & 640 & 1280 & 3.03M \\
        S6 & Combination & \eqref{eq:fusion_bl5} \& \eqref{eq:fusion_bl4} & 640 & 640 & 3.36M \\
        \bottomrule
    \end{tabular}
    \label{tab:models}
\end{table}

\begin{table*}[!htbp]
    \caption{The 200K training step WERs with different joint network structures.}
    \centering
    \begin{tabular}{lccccccccccc}
        \toprule
        ID  & de-DE & en-GB & en-US & es-ES & es-US & fr-FR & it-IT & ja-JP &  zh-TW & Avg. \\
        \midrule
        S0 \cite{Graves2013}{~~}   & 15.4 & 7.1 & 7.9 & 8.1 & 8.1 & 13.4 & 9.7 & 13.9 & 6.1 & 9.97 \\
        S1 \cite{Graves2013}{~~}   & 15.2 & 7.4 & 7.8 & 8.1 & 7.8 & 13.3 & 9.7 & 14.1 & 6.1 & 9.94 \\
        S2 \cite{Saon2021}    & 15.3 & 7.4 & 7.9 & 8.1 & 7.9 & 13.1 & 9.7 & 14.0 & 6.2 & 9.96 \\

        \midrule
        S3   & 15.1 & 7.4 & 7.6 & 7.9 & 8.0 & 12.9 & 9.4 & 13.7 & 6.1 & 9.78 \\
        S4   & 15.1 & 7.2 & 7.5 & 7.8 & 7.8 & 13.1 & 9.2 & 13.7 & 5.9 & \textbf{9.71} \\
        S5  & 15.0 & 7.2 & 7.5 & 7.7 & 7.6 & 13.3 & 9.1 & 13.6 & 6.0 & 9.67 \\
        S6   & 14.6  & 7.1  & 7.4 & 7.8 & 7.7 & 13.1 & 9.4 & 13.6 & 6.0 & \textbf{9.63} \\
        \bottomrule
    \end{tabular}
    \label{tab:results1}
\end{table*}
\vspace{-3mm}
\begin{table*}[!htbp]
    \caption{Final 800K training step WERs. Superscript ``${\text{reg}}$'' means to use regularisation with $m_1=$25K and $m_2=$200K.}
    \centering
    \begin{tabular}{lcccccccccccc}
        \toprule
        ID &  de-DE & en-GB & en-US & es-ES & es-US & fr-FR & it-IT & ja-JP & zh-TW & Avg. \\
        \midrule
        S0  & 14.2 & 6.6 & 6.8 & 7.5 & 7.5 & 13.0 & 8.6 & 12.9 &  5.6& 9.19 \\
        S0$^{\text{reg}}$  & 13.9 & 6.5 & 6.4 & 7.3 & 7.4 & 12.7 & 8.1 & 12.5 & 5.6 & \textbf{8.93} \\ 
        S1  & 14.3 & 6.6 & 7.0 & 7.5 & 7.6 & 13.0 & 8.6 & 12.6 & 5.6 & 9.20 \\
        S1$^{\text{reg}}$  & 14.0 & 6.5 & 6.8 & 7.2 & 7.5 & 12.7 & 8.3 & 12.4 & 5.4 & \textbf{8.98} \\
        \midrule
        S4  & 13.8 & 6.4 & 6.7 & 7.2 & 7.7 & 12.8 & 8.1 & 12.3 & 5.4 &  8.93 \\
        S4$^{\text{reg}}$  & 13.7 & 6.4 & 6.6 & 7.0 & 7.3 & 12.7 & 7.9 & 12.2 & 5.4 & \textbf{8.80} \\
        S5 & 13.8 & 6.4 & 6.6 & 7.1 & 7.4 & 12.8 & 8.2 & 12.3 & 5.5 & 8.90 \\
        S6  & 13.8 & 6.2 & 6.6 & 7.1 & 7.4 & 12.7 & 8.2 & 12.3 & 5.4 & 8.86 \\
        S6$^{\text{reg}}$  & 13.5 & 6.3 & 6.5 & 7.0 & 7.2 & 12.6 & 8.0 & 12.1 & 5.3 & \textbf{8.72} \\
        \bottomrule
    \end{tabular}
    \label{tab:results3}
\end{table*}
\vspace{-3mm}

\subsection{On joint network structures}
\label{ssec:results1}

The results of RNN-T models with different joint networks are presented in Table~\ref{tab:results1}. First, S0 and S2 result in similar WERs, which validates the finding in \cite{Saon2021}. Next, by increasing $D^{\text{joint}}$ from $640$ (S0) to $790$ (S1), the averaged WER was only slightly reduced by 0.03\%. We also tried to add more FC layers to the joint network that resulted in worse training loss values and higher WERs. 

Both gating (S3) and bilinear pooling (S4) outperformed FC fusion systems S0 and S1, where S1 has the same amount of parameters as S3 and S4 indicating that the lower WERs of S3 and S4 do not come from the extra model parameters. Furthermore, S4 outperformed S3 meaning that bilinear pooling can produce more expressive joint representations. Another advantage of bilinear pooling is having the flexibility to control $D^{\text{joint}}$ and $D^{\text{rank}}$ separately. Increasing $D^{\text{rank}}$ from 640 to 1280 allows S5 to have more parameters and lower WERs without increasing the input size of the big output layer with 16,384 nodes. At last, by stacking gating and bilinear pooling in the joint network, S6 leverages the complementarity of both structures and outperforms any other systems. In particular, with the same amount of parameters, S6 outperformed S1 by a 0.31\% WER reduction. In conclusion, all of our proposed systems (S3 -- S6) improved WERs considerably by increasing only a small percent of the model parameters (S0 has 144M parameters as given in Sec.~\ref{ssec:setup2}).


\begin{figure}[!htbp]
    \centering
    \includegraphics[width=1.0\linewidth]{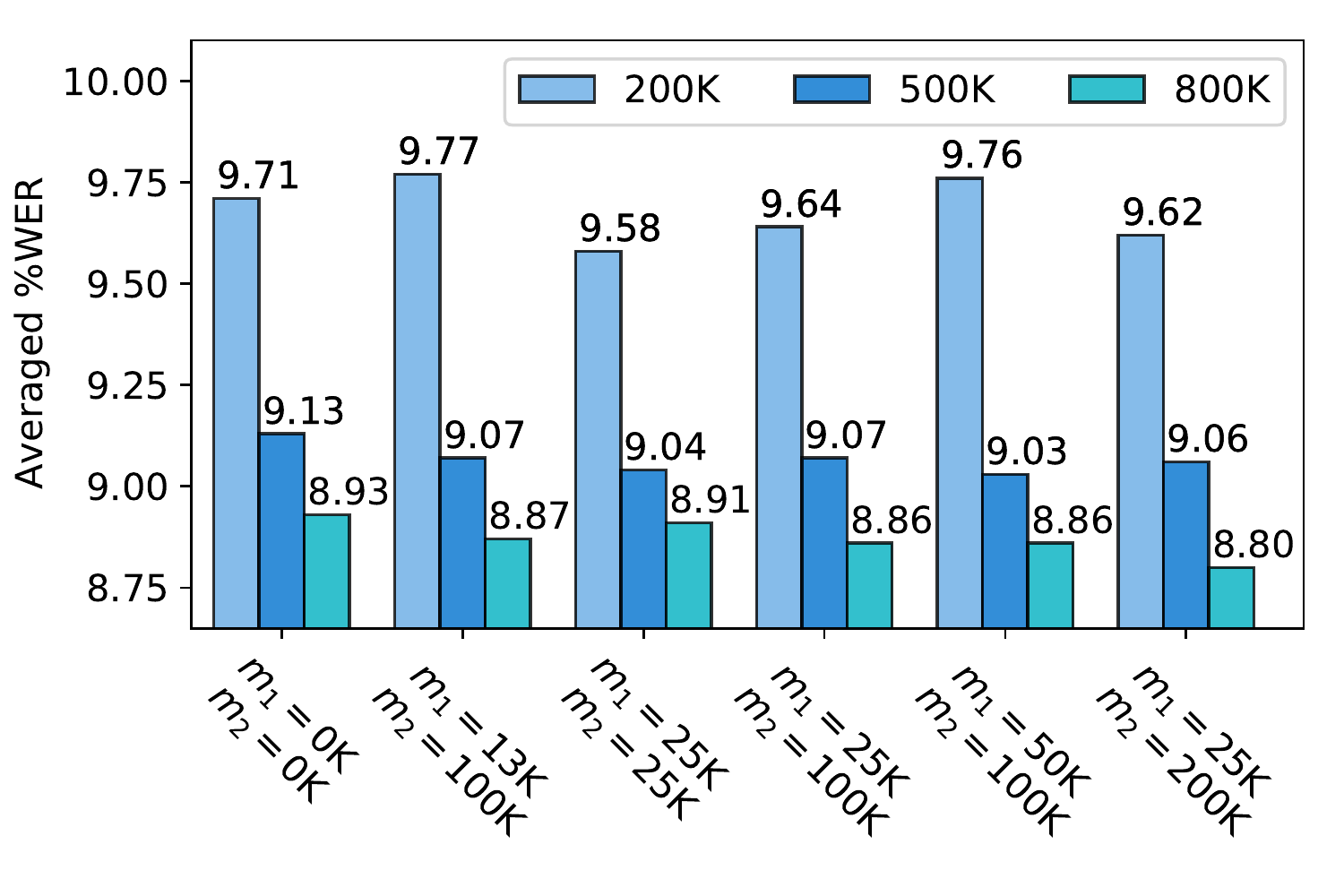}
    \vspace{-8mm}
    \caption{S4 system WERs with different hyper-parameters for prediction network regularsation.}
    \label{fig:regularisor}
\end{figure}

\subsection{On prediction network regularisation}
\label{ssec:results2}
The results of using the prediction network regularsation defined in Eqns.~\eqref{eq:regu1} and \eqref{eq:regu2} with different $m_1$ and $m_2$ values are shown in Fig~\ref{fig:regularisor}. S4, the bilinear pooling joint network with $D^{\text{rank}}=640$, is used in this section. The best results were found with $m_1=$25K and $m_2=$200K, which improved the averaged WER by 0.09\%, 0.07\%, and 0.13\% absolute with 200K, 500K, and 800K steps separately.



\subsection{Final results}
\label{ssec:results3}
The models are trained for 800K training steps and the final results are shown in Table~\ref{tab:results3}. Comparing S4, S5, and S6 to S1, better joint network structures lead to lower WERs without requiring more parameters. 
The prediction network regularisation improved averaged WERs by 0.26\%, 0.13\% and 0.14\% for S0, S4, and S6, although the improvements are not always consistent regarding each individual language. Compared to baseline system S0, our best-performing systems S4 and S6 with the regularisation method achieved 4.2\% and 5.1\% relative reductions in averaged WER by increasing only 1.15M and 2.63M parameters.

\section{Conclusions}
\label{sec:conclusions}
By viewing the function of the joint network of RNN-T as to fuse the acoustic and text representations derived from the acoustic encoder and prediction network, we propose in this paper to apply the structures widely used for multimodal representation fusion, including gating and bilinear pooling, to improve the joint network implementation. These structures are modified and combined to fit into RNN-T. A novel prediction network regularisation method is also proposed to make the fusion between acoustic and text representations more balanced. 
When evaluated using a large-scale multilingual voice search setup, 4.2\% and 5.1\% relative WER reductions were found by using the bilinear pooling and the combination structure separately along with the regularisation.


\vfill\pagebreak

\section{REFERENCES}
\label{sec:refs}
\renewcommand{\section}[2]{}
\renewcommand*{\bibfont}{\footnotesize}
\setlength{\bibsep}{3pt plus 0.3ex}
\bibliographystyle{IEEEbib}
\bibliography{refs}

\end{document}